\documentclass[
]{ceurart}

\usepackage{listings}
\usepackage{subcaption}
\usepackage{flushend}
\usepackage{microtype}
\usepackage{ccicons}
\usepackage{indentfirst}
\usepackage{hyperref}
\usepackage[dvipsnames]{xcolor}

\newcommand{\wims}{visualizations in motion}
\newcommand{\wime}{\emph{visualization in motion}}
\newcommand{\wimse}{\emph{Visualizations in motion}}

\ifpdf%                                % if we use pdflatex
  \pdfoutput=1\relax                   % create PDFs from pdfLaTeX
  \usepackage{graphicx}                % allow us to embed graphics files
  \DeclareGraphicsExtensions{.pdf,.png,.jpg,.jpeg} % for pdflatex we expect .pdf, .png, or .jpg files
\else%                                 % else we use pure latex
  \usepackage{graphicx}                % allow us to embed graphics files
  \DeclareGraphicsExtensions{.eps}     % for pure latex we expect eps files
\fi%

\graphicspath{{figures/}{pictures/}{images/}{./}} % where to search for the images

\newcommand{\inlinevis}[3]{\raisebox{#1}[0pt][0pt]{\includegraphics[height=#2]{#3}}}

\newlength{\pictureheight}
\newlength{\picturesep}
\newcommand{\bpstart}[1]{\vspace{1mm} \noindent{\textbf{#1.}}}

\begin{document}

\copyrightyear{2022}
\copyrightclause{Copyright for this paper by its authors.
  Use permitted under Creative Commons License Attribution 4.0
  International (CC BY 4.0).}

% \conference{MobileHCI'22: Workshop on New Trends in HCI and Sports,
%   September 28 -- October 1, 2022, Vancouver, Canada}
\conference{NTSPORT'22: New Trends in HCI and Sports Workshop at MobileHCI'22, October 1, 2022}

\title{Reflections on Visualization in Motion \\for Fitness Trackers}

\author[1]{Alaul Islam}[%
orcid=0000-0001-6900-3822,
email=mohammad-alaul.islam@inria.fr,
]
\fnmark[1]

\author[1]{Lijie Yao}[%
orcid=0000-0002-4208-5140,
email=lijie.yao@inria.fr,
]
\cormark[1]
\fnmark[1]
\address[1]{Universit{\'e} Paris-Saclay, CNRS, Inria, LISN, 91190, Gif-sur-Yvette, France.}

\author[1]{Anastasia Bezerianos}[%
orcid=0000-0002-7142-2548,
email=anastasia.bezerianos@lri.fr,
]

\author[2]{Tanja Blascheck}[%
orcid=0000-0003-4002-4499,
email=tanja.blascheck@vis.uni-stuttgart.de,
]
\address[2]{University of Stuttgart, Stuttgart, Germany}

\author[1]{Tingying He}[%
orcid=0000-0002-9670-5587,
email=tingying.he@inria.fr,
]

\author[3]{Bongshin Lee}[%
orcid=0000-0002-4217-627X,
email=bongshin@microsoft.com,
]
\address[3]{Microsoft Research, Redmond, WA, USA}

\author[4]{Romain Vuillemot}[%
orcid=0000-0003-1447-6926,
email=romain.vuillemot@ec-lyon.fr,
]
\address[4]{Universit{\'e} de Lyon, {\'E}cole Centrale de Lyon, CNRS, UMR5205, LIRIS, F-69134, France}

\author[1]{Petra Isenberg}[%
orcid=0000-0002-2948-6417,
email=petra.isenberg@inria.fr,
]

\cortext[1]{Corresponding author.}
\fntext[1]{These authors contributed equally.}

\begin{abstract}
 In this paper, we reflect on our past work towards understanding how to design visualizations for fitness trackers that are used in motion. We have coined the term ``visualization in motion'' for visualizations that are used in the presence of relative motion between a viewer and the visualization. Here, we describe how visualization in motion is relevant to sports scenarios. We also provide new data on current smartwatch visualizations for sports and discuss future challenges for visualizations in motion for fitness trackers.
\end{abstract}


\maketitle

%======================================

\section{Introduction}
Fitness trackers, such as smartwatches and fitness bands record a variety of data. Most of these devices also visualize the collected data and make it immediately available to wearers. Smartwatch faces, in particular, have become mini data dashboards that can give an overview of data such as step counts, heart rates, locations, sleep information or even device-external data such as the current temperature or weather predictions. Due to their small screen size and usage context, fitness tracker screens pose several novel and interesting challenges to visualization: visualizations need not only to be small and glanceable but also often to be read in motion. For example, when an athlete trains for a race, they can only afford quick glances at a smartwatch while running, to concentrate on the path to take and avoid accidents. As stopping the race to look at a watch is not a desired option, the watch needs to be read while the runner's body, including their arms, is moving. During a quick glance at the tracker, the athlete may want to take in multiple information at once: current race time, heart rate, or distance run are just three examples. Unfortunately, there is still little advice on how to design effective information dashboards for fitness trackers, and existing designs are built without strong empirical foundations.

To address this problem we have recently begun to work on two directions: a) visualizations in motion, in which we assess the effects of motion on the perception of visualizations and b) visualizations for fitness trackers and, in particular, smartwatches. In this paper, we briefly introduce our past work with a focus on smartwatch-type fitness trackers, provide some new data on existing smartwatch faces for sports, and outline dedicated challenges for visualization in motion for fitness trackers. 

\section{Background}
While neither visualization in motion nor fitness tracker visualization has a long history of research, some relevant past work does exist. The following background presents the definition of visualization in motion and briefly outlines the larger research space. The second section focuses on fitness tracker visualization in the context of health and sports, and that of visualization in motion in relation to fitness trackers.

\subsection{What is Visualization in Motion?}
In our recent paper \cite{Yao:2022:VisInMotion}, we defined \wime\ as:

\begin{quote}
    \emph{Visual data representations that are used in contexts that exhibit relative motion between a viewer and an entire visualization. }
\end{quote}

\wimse\ specifically concern relative movement between visualizations and viewers and, therefore, they are different from animation of visualization components that are meant to express highlights, to smooth transitions between views \cite{StaggeringAnimation:2014, GraphDiaries:2014, SmoothViewTransitions:2007, MotionAnimatedScatterplots:2019}, or to morph between different representations \cite{StagedAnimation:2007, RollingDice:2008, ConeTrees:1991, 3DTreemaps:2005, VisualLiteracy:2015}. Relative motion between entire visualizations and viewers is relatively common in the sports context, but has not been explored in depth. Examples include stationary players who sit in front of a screen while playing a sports game, in which a game character (e.g., an American football player) moves with attached donut charts showing data related to the character (\autoref{fig:game}), audiences sitting in a stadium while watching an augmented basketball game that shows data next to players (\autoref{fig:sports}), people walking across or driving by physicalizations (Figure \ref{fig:walkablevis} and \ref{fig:newhaven}), a person reading how many calories they have burned from a fitness tracker while exercising (\autoref{fig:smartwatchmotion}), and a traveler navigating using a phone while walking (\autoref{fig:mobilevis}). These scenarios can be grouped into three categories of \wime: 

\begin{itemize}
    \item moving visualizations \& stationary viewer
    \item stationary visualizations \& moving viewer
    \item moving visualization \& moving viewer
\end{itemize}

In this paper, we bring together our work on fitness trackers and visualization in motion, and thus focus on the last group that involves moving visualizations \& moving viewer. Our specific focus on fitness trackers was motivated by the fact that they already carry visualizations and their wearers' are not only often moving but also have information needs while on the go, such as learning about their performance and condition.

\setlength{\pictureheight}{3.23cm}
\setlength{\picturesep}{1mm}

\begin{figure*}
    \centering
   \subcaptionbox*{\scriptsize{Moving visualization \& stationary viewer.}}%
                  {\includegraphics[height=\pictureheight]{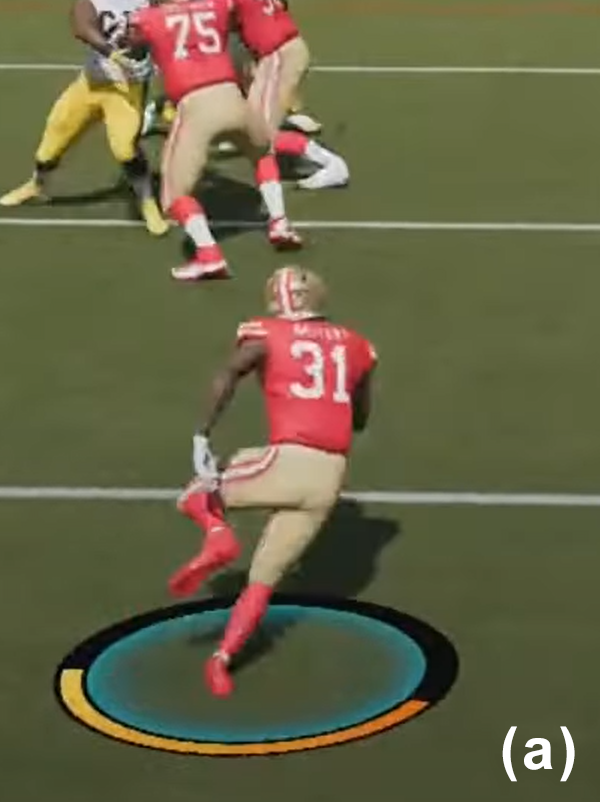}
                  \phantomsubcaption\label{fig:game}
                  \includegraphics[height=\pictureheight]{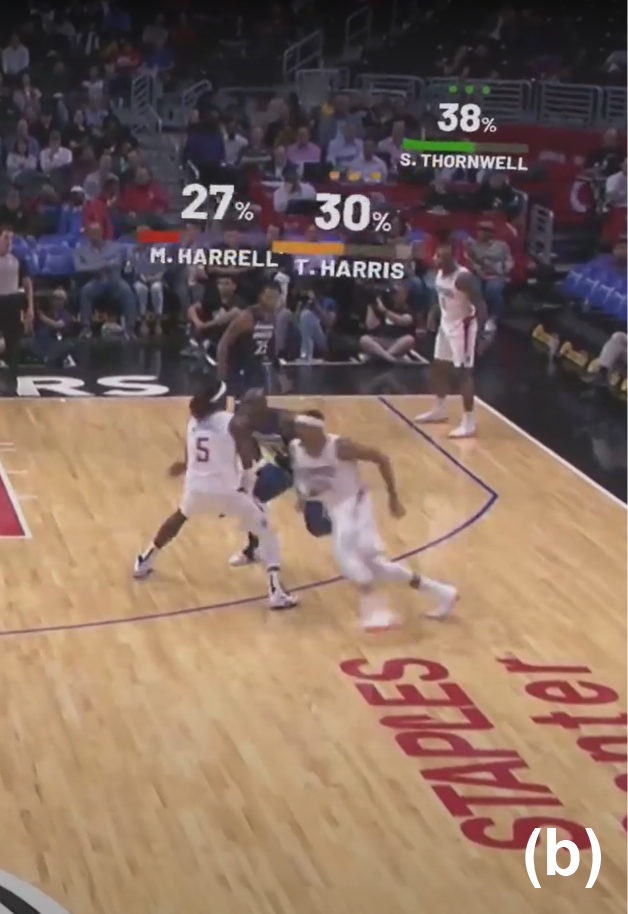}
                  \phantomsubcaption\label{fig:sports}
                  }%
    \hspace{0.1 cm}
    \subcaptionbox*{\scriptsize{Stationary visualization \& moving viewer.}}%
                  {\includegraphics[height=\pictureheight]{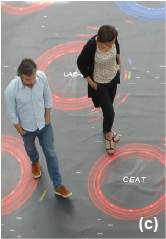}
                  \phantomsubcaption\label{fig:walkablevis}
                  \includegraphics[height=\pictureheight]{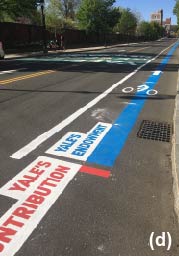}
                  \phantomsubcaption\label{fig:newhaven}
                  }%
    \hspace{0.1 cm}
    \subcaptionbox*{\scriptsize{Moving visualization \& moving viewer.}}%
                  {\includegraphics[height=\pictureheight]{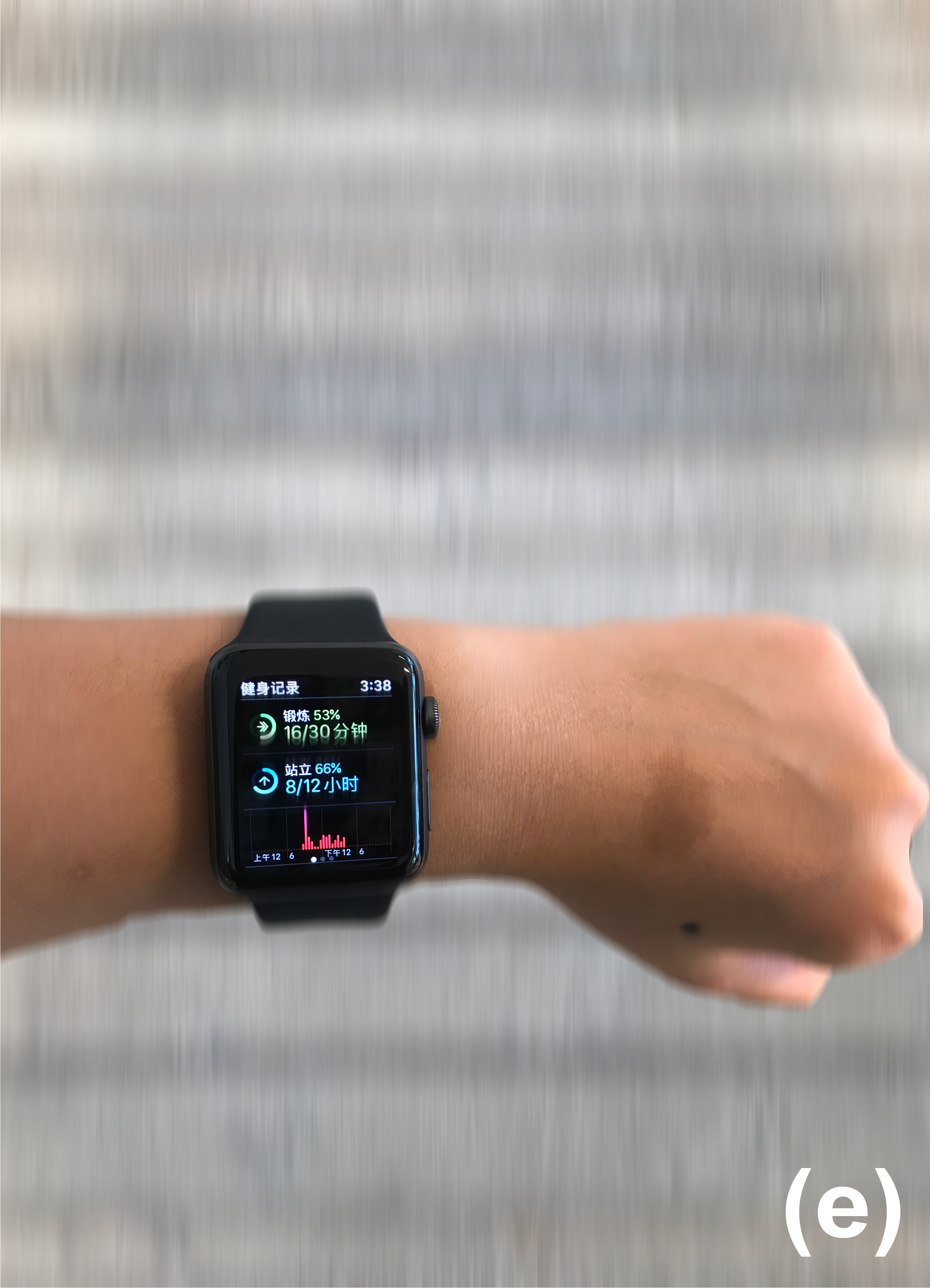}
                  \phantomsubcaption\label{fig:smartwatchmotion}
                  \includegraphics[height=\pictureheight]{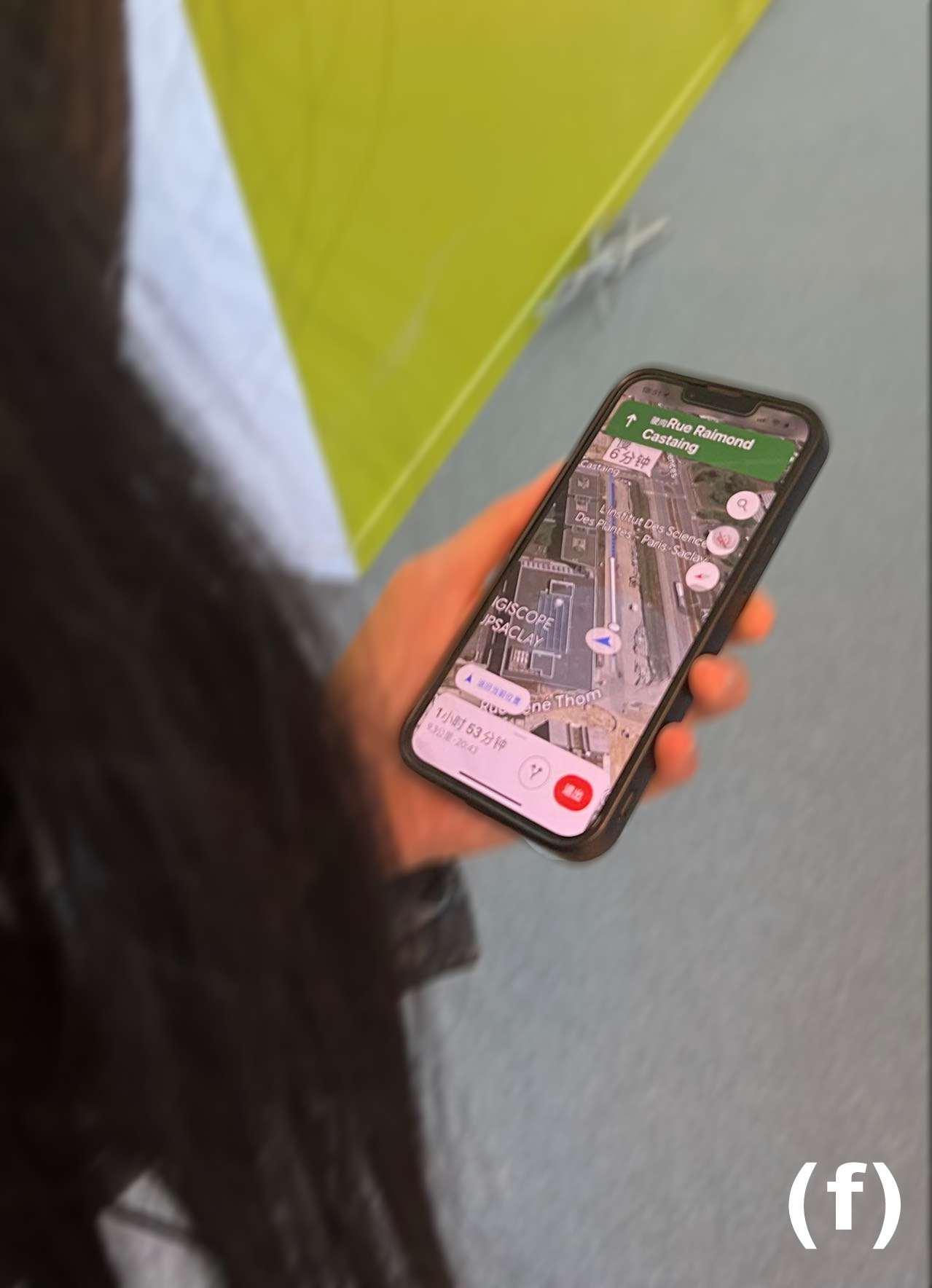}
                  \phantomsubcaption\label{fig:mobilevis}
                  }%
    \caption{\protect Visualization scenarios that involve different types of relative movement between viewers and visualization. \emph{Image permissions are listed in the acknowledgments.}}
  \label{fig:teaser}
\end{figure*}

\subsection{Fitness Tracker Visualizations}
Choosing what type of data and how to show it to wearers is a fundamental challenge that can impact how devices are adopted. In our own work, we used commercial fitness trackers such as fitness bands and smartwatches because our focus was on data representation and not on the development of new technologies. However, we acknowledge that many types of wearable displays have been proposed \cite{Heller:2021:DesignSpaceWearableDisplays} and discuss some challenges related to these in our research agenda. 

Niess et al.~\cite{Niess:2020:Fitness-Tracker-Visualisations} studied the impact of various approaches to represent unmet fitness tracker goals through visualization on rumination, highlighting that multicolored charts on fitness trackers may lead to demotivation and negative thought cycles. Havlucu et al.~\cite{Havlucu:2017:Tennis-Players} interviewed 20 professional tennis players and found that the players' abandonment of their trackers was due to the type of information displayed on the fitness trackers. Participants wished to see tennis-specific data, recovery rate, and nutrition, as well as precise technical data regarding their tennis performance, such as where the ball hit the racket, the speed of a stroke, how the ball bounced off the floor, general mobility on the court, as well as weak points and errors regarding their own game.

Outside of the professional sports context, smartwatches also have a lot of potential to be an essential part of the personal health movement. Yet, even with a potentially large target audience, visualization guidelines for fitness trackers are still sparse. Most of the past studies discussed health and physical activity data representations on smartwatches and mentioned the challenges of representing these data types~\cite{Neshati:2019:Challenges-Displaying-Health-Data}. Van Rossum \cite{van:2020:smartwatch-activity-coach} suggested smartwatch visualizations aiming for easy-to-understand, clear visuals, using a black background for contrast, and less disturbance in dim environments. Albers et al.~\cite{Albers:2014:Task-Driven-Eval} showed that the tasks that wearers do when exploring a visualization are influenced by the visualization's design and choices of visual factors (e.g., position, color), mapping variables (e.g., raw data, averages), and computational variables (how aggregated data are computed). Pekta{\c{s}} et al.~\cite{pektacs:2021:Smartwatch-Diabetes-Diary-Application} showed how visualization using icons and emoji on warnings and alerts could motivate wearers to monitor health related information. In contrast to these works, we are interested in fitness tracker visualization in motion, specifically when wearers are on the move during sports activities, which is less explored in the literature.

\subsection{Visualization in Motion for Wearable and Mobile Devices}
In the context of wearable devices, relative motion is most often created when both viewers and visualizations are in motion such as during a run or walk. Several previous studies on mobile phones have shown that walking increased workload and reduced performance in reading tasks \cite{Schildbach:2010:ReadingWhileWalking,Mustonen:2004:TextLegibilityWhileWalking,Vadas:2006:ReadingOnTheGo}. As cognitive resources need to be similarly shared between navigation and reading data, it seems reasonable to expect similar negative effects for \wims\ on fitness trackers. However, the exact effects have not been studied in enough depth to make recommendations for the design of visualization in motion. Although moving participants were involved in the studies by Schiewe et al. \cite{SportsActivitiesonSmartwatches} on visualizations for real-time feedback during running activities, by Amini et al. \cite{Amini:2017:DRF} on in-situ health and fitness data exploration for fitness trackers, and by Langer et al.~\cite{Langer:2021:Mountain-Biking-Smart-Wearables} on crash risk indication applications for sports smartwatches in the context of mountain biking, the effects of relative movement between displaying charts and exercising people received little to no dedicated attention.

However, we may take inspiration from another research area containing moving viewers and moving visual targets: immersive analytics. Literature from psychology has shown that walking in VR may have negative impact on multi-object tracking \cite{Laura:2010:SelfmotionMultipleObjectTracking}. In fact, several research efforts in VR have targeted a viewer's motion, such as examples illustrated in Locomotion Vault \cite{Di:2021:Locomotion}. Examples collected by Locomotion Vault includes one showing that in a virtual environment, the viewer's spatial memory can benefit from common motion effects such as walking. Grioui and Blascheck \cite{Grioui:2021:HeartRateVirtualSmartwatch} conducted a first pilot on heart rate reading from a virtual smartwatch in the context of a VR game that gave preliminary indications that heart rate visualizations in the form of summary charts might be effective for making decisions to reach heart rate goals. Thus, how people will perform when reading visualizations in motion in an immersive environment still requires more dedicated work.

In summary, literature on how visualizations are read under motion or how they should be designed to be effective in a sports context is still too sparse to make clear recommendation. Next, we outline some of our past research on how visualizations are currently designed for sports-related smartwatch faces before moving on to recommend a research agenda for this emerging topic.

\subsection{Guidelines for Visualization Design}
Ample evidence exists that visualization choice and design will impact the readability of visualizations without motion. These design factors need to be explored again specifically for micro (very small) visualizations on fitness trackers used while in motion. Example factors include representation type \cite{ZimekiewiczKosara1, ZimekiewiczKosara2}, the visualization complexity \cite{Cleveland1984GraphicalPT, FourExperimentBars:6876021}, the decoration of the representation \cite{ImprovingPerceptionAccuracyinBarCharts:2018, ReadabilityPictorialBarChart:2017, DecoratedBarChart:2015}, the size of the visualization \cite{doughnutProportion:2018}, the color selection \cite{Szafir:2018:ColorDifference, Zhou:2016:ColormapsSuvery}, and specifically for a micro display with limited space, the visualization density \cite{Neshati:2021:SpaceDensitySmartwatch}. 
Previous research has shown that cognitive overload can occur when too much information is presented during attention-demanding sports like tennis~\cite{Hayati:2019:Activity-Trackers-for-Performance-Sports}. As such, information needs to likely be minimal, context-specific, and glanceable to the wearers. 
Gouveia et al.~\cite{Gouveia:2015:Activity-Trackers-Engagement} showed that the average wearer's involvement with the trackers was brief, 5-sec, without further interaction. However, we expect the duration to be much shorter than that during sports activities. Previous research showed that people could effectively read even complex sleep visualizations on fitness trackers~\cite{Islam:2022:Sleep-Data-Visualizations}, perform simple comparison tasks with visualizations on smartwatches within several hundred milliseconds~\cite{SmartWatch:Tanja}, providing evidence that visualizations could be effective forms of data representations in the context of fitness trackers.
However, ambient illumination, lighting effects~\cite{Kerber:2017:Ambient-Illumination, Colley:2016:Informative-Illumination-Snowboard} or motion textures~\cite{Lockyer:2011:Motion-Textures} for fitness trackers could also be a possible way to achieve glanceability during attention-requiring sports activities, during which wearers may get feedback through  color changes, brightness levels, or texture changes. 

Yet, what exact limits are for how much information to be displayed and at what sizes is still underexplored.  Should all data be represented with a visualization? If not, what would be a good number to have? 

\section{Current Visualizations for Smartwatch Faces}

The screens that wearers of smartwatches look at most often are the ``home'' screen or ``smartwatch face.'' These smartwatch faces show time but also a variety of additional data to wearers and are often designed for specific themes, including sports. To better understand what type of data current sports watch faces show to wearers and how this data is represented, we conducted a systematic review of sports category tagged watch faces from the Facer App~\cite{Facer:2014}.

\subsection{Data Collection}
We decided to collect watch faces from the Facer App, one of the most popular smartwatch face distribution websites. It contains a Top100 page that lists the premium or free watch faces of Apple and WearOS/Samsung smartwatches. Because the list for the Apple Watch did not consistently contain 100 faces, we chose to focus on the WearOS/Samsung watch faces. Nevertheless, the watch faces we have collected can also be used in “square-shaped” Apple watch-like watches. We manually collected the metadata of the top 100 smartwatch faces every Sunday for one month because the premium list was recalculated on Sunday, starting from March 14, 2021. The metadata collected for each watch face included its rank, name, category,  link, and thumbnail. Among the 400 top watch faces we collected, 184 were unique watch faces, as several appeared in the top 100 for multiple weeks. From the 184 unique smartwatch faces, we found that 42 watch faces were categorized as sports watch faces.

We analyzed the watch faces with the extracted image. If a design was unclear from the thumbnail image, we went to the Facer website to look at the simulated watch face graphic. We group our results according to the data shown on a sport-tagged watch face and data representation types.

\subsection{What Data is Shown on Sport Watch Faces?}
One of the difficulties with designing data visualizations for smartwatch faces is that these visualizations typically show many types of independent data (steps, weather, battery levels, etc.) that need to be shown in a coherent watch face design. These non-time/date data functionalities on smartwatches are called \emph{complications}  \cite{Jackson:2019:Design-Fundamentals}. In this sense, watch faces with several complications can be considered small personal dashboards with distinctive design challenges. These design challenges include limited display space for a large number of possible complications, device form factors, as well as the specific context of use, in our case, sports activity, that often requires information to be readable at a glance. In addition, watch faces require that time or date is readable and often remains the primary data shown. We present our findings from analyzing 42 sports watch faces in the following.

\textbf{Number of Data Types.} The watch faces from the sports category contained a median of six data types, similar to Islam et al.’s smartwatch face survey~\cite{Islam:2020:Smartwatch-Survey}, in which participants reported a median of five data types.

\textbf{Types of Data.} Health \& fitness related data were the most common. We found 41.05\% health \& fitness related data, among which step count and heart rate were the most common. However, the watch faces also contained 35.37\% weather \& planetary data such as temperature and sky condition, and 23.58\% device \& location related data such as watch and phone battery level. Among the top 10 most common data items, we found four (step count, heart rate, distance traveled, and calories burned) that were health \& fitness related data. The day's temperature, including weather information, moon phase, and sunset/sunrise time, were the most frequently displayed weather data on sports watch faces. Watch battery level, which is the first and most displayed data item, as well as phone battery level were device location related data items displayed on the sports watch faces.

\subsection{How are the Sport Watch Faces Designed?}

\begin{figure}[!t]
    \centering
    \includegraphics[width=.8\textwidth]{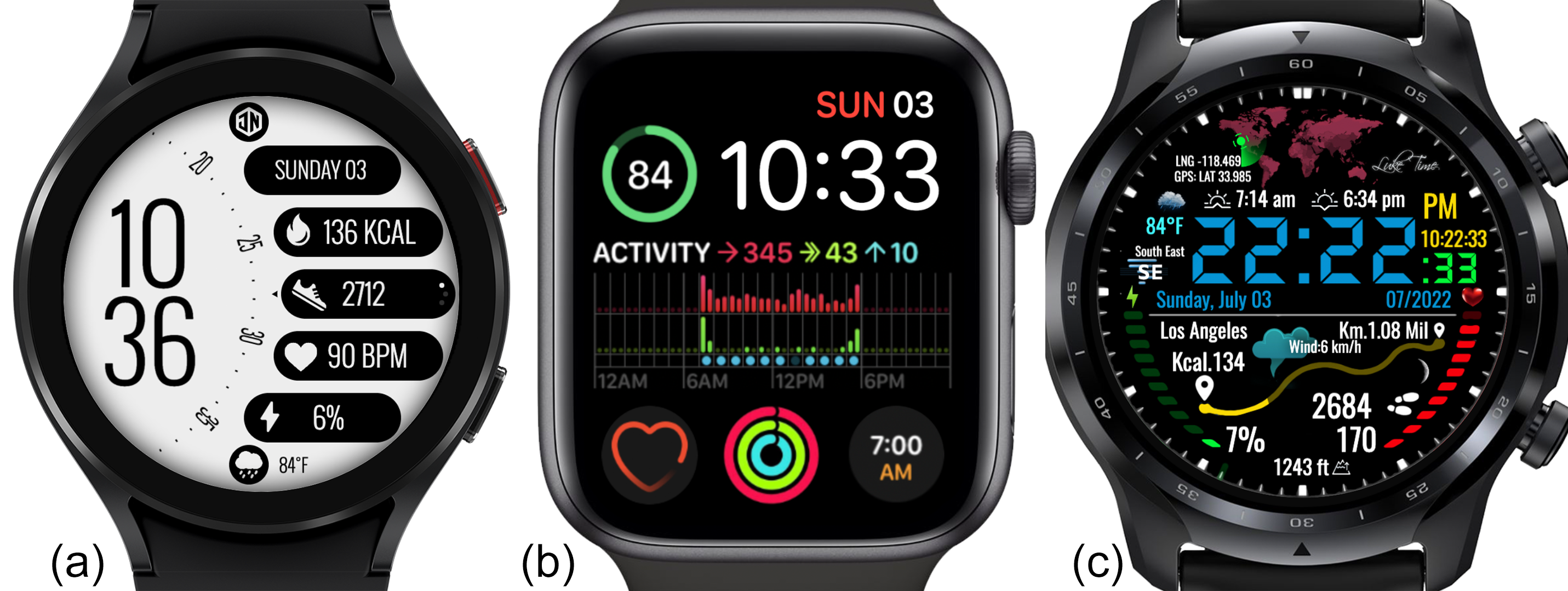}
    \caption{Example of sports smartwatch faces. Watch face design by: a) JN--Rolling (JN--WatchFaces), b) New Design (Enid Rodriguez), and c) Voyager GPS-01RB Sport 24H (Luke Time) from Facer App~\cite{Facer:2014}.}
     \label{fig:example_sports_watch_faces}
\end{figure}

Watch faces generally were comprised of components that we group into those representing time, complications, and decorations, each with its representation styles. Some of the example sports smartwatch faces are shown in~\autoref{fig:example_sports_watch_faces}.

\bpstart{Time display} Watch faces can be divided into digital, analog, and hybrid watch faces depending on the time display. Digital watch faces represent time information as HH:MM:SS for hours, minutes, and potentially seconds. Analog watch faces typically use the hour, minute, and second hands to indicate the time, to resemble conventional analog watches. Hybrid watch faces show both digital and analog time displays. Our analysis showed that the majority of premium sports watch faces were hybrid watch faces (40.5\%), followed by digital watch faces (33.3\%) and analog watch faces (26.2\%).

\begin{figure}[!t]
    \centering
    \includegraphics[width=.8\textwidth]{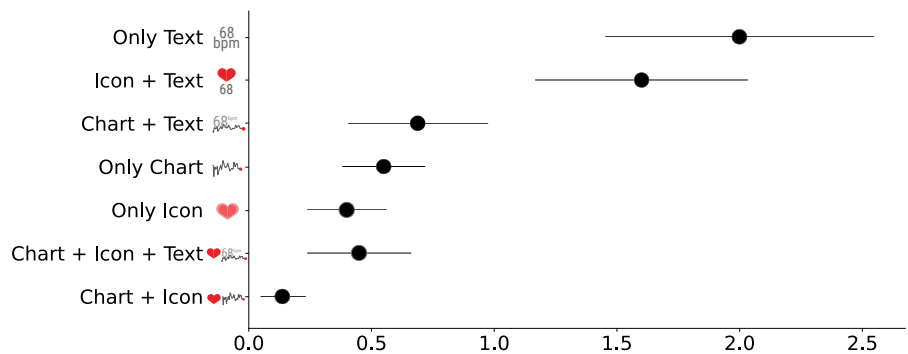}
    \caption{The average number of representation types on each sports watch face.}
     \label{fig:average_data_representation_type}
\end{figure}

\bpstart{Data Types Representations} We found seven ways of representing data, that were based on combinations of text, icons, and charts, as
shown in~\autoref{fig:average_data_representation_type}. As icons, we classified graphical content not in the strict semiotic sense and more analogously to how they were used in computing. Here icons are a type of image that represents something else. As such our icons can be both semiotic symbols \inlinevis{-2pt}{1.2em}{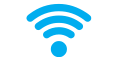} or icons \inlinevis{-2pt}{1.2em}{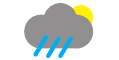}. \autoref{fig:average_data_representation_type} shows the average number of representation types on each sports watch face. A simple text label (Only Text, \inlinevis{-2pt}{1.2em}{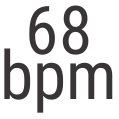}) was the most common representation type and was used for 2 data types on average on each watch face (\textit{M} = 2, 95\% CI: [1.45, 2.57]). Icons accompanied by text labels (Icon+Text, \inlinevis{-2pt}{1.2em}{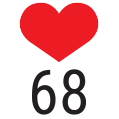}) were the second most common (\textit{M} = 1.6, 95\% CI: [1.17, 2.05]). In Islam et al.’s survey~\cite{Islam:2020:Smartwatch-Survey}, Icon+Text had been the most common representation type, used to display two kinds of data types on average on each watch face (\textit{M} = 2.05, 95\% CI: [1.78, 2.32]) followed by Text Only (\textit{M} = 1.38, 95\% CI: [1.13, 1.66]). Both evaluations clearly show that text is the most frequent way to represent data on watch faces whiles charts or charts combined with text or icons were rare in practice. Chart+Text \inlinevis{-2pt}{1.2em}{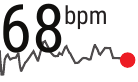} (\textit{M} = 0.69, 95\% CI: [0.4, 1]), Chart Only \inlinevis{-2pt}{1.2em}{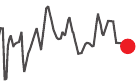} (\textit{M} =  0.55, 95\% CI: [0.38, 0.71]), Chart+Icon+Text \inlinevis{-2pt}{1.2em}{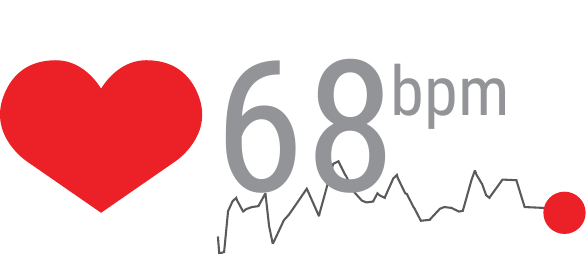} (\textit{M} = 0.45, 95\% CI: [0.24, 0.71]), and Chart+Icon \inlinevis{-2pt}{1.2em}{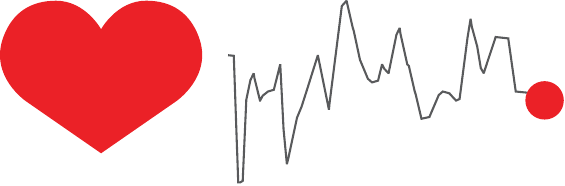} (\textit{M} = 0.14, 95\% CI: [0.05, 0.29]) appeared on average less than once per sports watch face. One notable difference in the data was the difference in Only Icon displays. Examples for representations that rely purely on a small image, such as weather icons (\inlinevis{-2pt}{1.2em}{weather-icon}\inlinevis{-2pt}{1.2em}{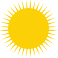}) are still rare on watch faces. In this sports watch faces analysis, Only Icon displays were, as expected, much more rare and we saw them only for weather data (14\texttimes), moon phases (2\texttimes), and wind directions (1\texttimes).

\section{Research Agenda for Visualizations in Motion on Fitness Trackers}
When fitness trackers are worn during sports activities that involve moving one's arms (walking, running, swimming, skiing, climbing etc.), the displays will be in motion relative to the wearers gaze. Depending on the activity the relative motion will be more or less predictable, and more or less quick, and the wearer will have different information needs. Next, we outline several aspects of visualization design for fitness trackers that require more research when the intended use involves motion.

\subsection{Understanding the Influence of Motion}
Motion characteristics such as speed, acceleration, trajectories, or direction may have an impact on the readability of visualizations. Yao et al. \cite{Yao:2022:VisInMotion} conducted two first evaluations about how donut charts' moving speed and trajectory affected the reading accuracy. Their results showed that participants' performance was better on linear trajectories and slow speed than that on irregular trajectories and fast speed. However, in their experiment, all participants were stationary and sat in front of a screen larger than 13 inches. Because fitness trackers have a much smaller display size and many application scenarios involve moving viewers, the impact of motion characteristics require further research in this context.  

In addition, motion in realistic indoor and outdoor scenarios will entail additional challenges such as changing lighting conditions, the presence of equipment, and a primary task. The type of sport itself will largely determine the types of motion characteristics and the extent of secondary factors. As such dedicated research is likely necessary. The characteristics of the different sport types determine the continuity of the viewer's movement and the presence of required sports equipment can directly affect the viewer's ability to read or even attach a fitness tracker. For example, swimming goggles may filter certain light, reduce the field of view, or having to wear heavy coats while skiing might make it difficult to access a wrist-worn smartwatch screen. Finally, the needed concentration on primary tasks determines the length of time available for the viewer to read from their fitness tracker.

\subsection{Understanding How Context Matters}
The primarily intended context of a fitness tracker's use needs to be considered in its graphical and interaction design.  The default for some Garmin watches, for example, is to show data during exercise using a large black font on a white background. No visualizations are shown. Is this the most effective way to communicate data to wearers or the one that ensures the most safety during other primary tasks? Especially contexts with divided attention, for example, glancing during driving, cycling, or running, require further research attention. Here, viewers can only afford quick glances at watch faces. Visualizations in these settings are difficult to evaluate and test, and future work is needed not only on which visualizations are glanceable but also on study methodologies to actually measure glanceability during sports activities.

Another important factor is the intended task context for watch faces. Islam et al.~\cite{Islam:2022:context-specific-smartwatch-vis} showed that with dedicated ideation exercises, watch faces could be easily envisioned that target specific usage contexts such as sightseeing in their case. Digital watch faces are easy to switch, and it would be interesting to study the impact of dedicated but changing watch faces on wearers. As mentioned above, the design as well as the placement of fitness trackers needs to likely be specific to different types of sports. During swimming, for example, it is almost impossible to read in-situ performance such as heart rate or lap times unless the swimmer stops to see their smartwatch. Completely new technology might be needed to support certain sports well. Attaching visualizations directly on the bottom of a swimming pool may, for example, be more effective information displays for swimmers than a wearable device.

\subsection{Display Types}
Being able to focus on the primary task is vital during sports activities. The capabilities of the technology chosen to display visualizations in motion may have a large impact on how well athletes can focus on their performance. Heller et al.\ \cite{Heller:2021:DesignSpaceWearableDisplays} discussed a design space for wearable displays with two main dimensions: on-body placement and display content. As they showed, branching out from commercial fitness trackers to wearable accessories, clothing, or skin and body projections is a possibility and ample research opportunities for visualization design exist---not only for performance-oriented displays but also for ambient visualization \cite{Genc:2019:TxtilePatterns}. In addition, interaction with these displays could be taken into account. Burstyn et al.~\cite{Burstyn:2015:DisplaySkin}, for example, presented an interactive wrist worn device prototype in which the display could adjust to the wearer's body pose. As hand and arm postures can change rapidly during an activity, fitness trackers that are body-pose aware could change the rotation, size, and location of a visualization to be most readable. While not technically ``visual,''  another possibility to represent data is to examine sonification, which involves mapping information to sound characteristics. Godbout and Boyd~\cite{Godbout2010CORRECTIVESF} showed how speed skaters are alerted with sonification when something is wrong and additionally how they are informed in which way they performed incorrectly. It would be valuable to explore further how to leverage sonification to facilitate more fluid smartwatch interaction while ``on the go.'' 
Apart from sonification, other non-visual channels such as touch can also be useful in eyes-free contexts. For example, in Neshati et al.'s work \cite{Neshati:2020:SkinDraggingVis} on tactile line chart reading,  a tip on participants' skin allowed them to perceive the data. Similarly other tactile methods, including vibration, should be further explored to determine what kind of and how well data can be read from this sensory channel.

\section{Conclusions}
The goal of this paper is to bring attention to an interesting and still wide-open area of research in the domain of sports: visualization in motion for fitness trackers. We explained \wime\ as a direction of research and how it is relevant to fitness trackers. We also provided evidence of current practice of sports watch faces and outlined in a brief research agenda what questions remain to be explored. 
Our survey on sports watch faces showed that wearers had six complications on average, in addition to time on their watch faces. The highest number of complications was 16. Future research is needed to determine how many complications on a small smartwatch display can effectively communicate information to wearers when doing sports activities. There are several avenues of scalability to explore: more data, smaller size, and more visualization in the context of smartwatch visualization in motion and specifically the glanceability of these visualizations.
A general research question that remains to be solved is how visualizations are read and studied in the context of real application scenarios. In summary, we discussed the need to research the following aspects of visualization design for fitness trackers:
\begin{itemize}
    \item readability of different visual designs such as chart types, color choices, etc.,
    \item scalability of visualization numbers, size, and types of data items,
    \item glanceability of different visual designs,
    \item the impact of motion factors in the context of specific sports,
    \item the impact of readability of visualizations under divided attention,
    \item and different sensory modalities for data representation on fitness trackers.
\end{itemize}

Visualizations in motion are, however, also relevant in other sports-related scenarios as we outlined earlier: augmented reality when watching sports, sports video games, or when static visualizations are read by athletes during their activities. Similar to the specific challenges outlined for fitness trackers, these other scenarios require future research. 
We hope that our paper will be used as a foundation for discussion and inspiration for future work that tackles the interesting remaining research questions on visualization in motion.

\begin{acknowledgments}
   This work was partly supported by the Agence Nationale de la Recherche (ANR), grant number are ANR-18-CE92-0059-01 and ANR-19-CE33-0012. Tanja Blascheck is funded by the Ministry of Science, Research and Art Baden-Württemberg.
   
   \emph{Image credits: All images copyright of the person granting permission: \autoref{fig:game}: non-commercial use under agreement \cite{EA:2022:UA}, \autoref{fig:sports}: SportBuzzBusiness \cite{SBB}, \autoref{fig:walkablevis}: Dario Rodighiero, \autoref{fig:newhaven}: Eddie Camp of \emph{Respect New Haven.}}
\end{acknowledgments}

\bibliography{lijie-alaul}

\appendix

\end{document}